\documentclass[prl, twocolumn]{revtex4-1}

\usepackage{amsmath}
\usepackage{amssymb}
\usepackage{graphicx}
\usepackage{wasysym}

\begin{document}

\title{Tunnel conductance spectroscopy via harmonic generation in a hybrid capacitor device}

\author{Ian Appelbaum}
\affiliation{Department of Physics and Center for Nanoscience and Advanced Materials, University of Maryland, College Park, Maryland 20742, USA}

\begin{abstract}
We address the measurement of density of states within and beyond the superconducting gap in tunnel-coupled finite-size nanostructures using a capacitive method. Third-harmonic generation is used to yield the full differential conductance spectrum without destruction of the low dimensionality otherwise induced by intimate ohmic coupling to an electrode. The method is particularly relevant to attempts to discern the presence of the fragile Majorana quasiparticle at the end of spin-orbit-coupled nanowires in appropriate magnetic field conditions by their signature mid-gap density of states.  
\end{abstract}

\maketitle

The recent observation of a magnetic field-induced zero-bias conductance (ZBC) feature in the proximity superconductivity tunneling spectrum gap of a semiconductor nanowire has spurred claims of a solid-state bound Majorana fermion.\cite{Mourik_Science2012, Das_NaturePhys2012, Deng_NanoLett2012} If true, this topological excitation may become the basis for realistic proposals of a fault-tolerant quantum computing scheme.\cite{Kitaev_Uspekhi2001, Read_PRB2000, Nayak_RMP2008} However, there still remain several intriguing inconsistencies in the available evidence before predictions\cite{Sau_PRL2010, Sengupta_PRB2001, Flensberg_PRB2010, Alicea_PRB2010, Wimmer_NJP2011, Stanescu_PRB2011, Law_PRL2009} are unambiguously confirmed.  

Perhaps most obvious is the failure of the superconducting gap to close before the new ZBC feature appears. This is particularly problematic because the gap closure marks a topologically critical boundary for the expected formation of the Majorana quasiparticle. Although recent theories have successfully argued that the discrepancy is caused by measurement of the wire-end \emph{local} density of states (DOS) and not inconsistent with the topological phase,\cite{Stanescu_PRL2012} a true measurement of ``global'' density of states gap collapse in magnetic field would provide much-needed confirmation and significantly settle debate.

Experimentally, this need to capture the global DOS of a one-dimensional nanowire presents a difficult measurement problem. Since the Majorana state is a property of the boundaries of the one-dimensional nanowire, we cannot simply contact it with a conductor to prepare a bulk tunnel conductance measurement and expect the dimensionality - and topological state - to persist.  

Here we propose a novel technique designed to solve this problem and measure the proximity superconducting gap of a one-dimensional nanowire without any perturbing ohmic contact. This approach, which exploits harmonic generation under AC excitation\cite{Shapiro_JAP1967}, is especially conducive to measuring the properties of ensembles of quantum wires lithographically- or gate-defined\cite{Reuther_arxiv2013} from planar epitaxial heterostructures, and can be performed in a simple two-terminal geometry. To substantiate the proposal, a numerical simulation of harmonic generation using the discrete 1-d wire spectrum as a function of magnetic field is performed. Finally, a proof-of-principle experiment using a thin-film superconducting-normal metal tunnel junction is suggested to provide a means to benchmark against a trivial ``hard gap'' system. 

\begin{figure}
\centering
\includegraphics[width=2.75in]{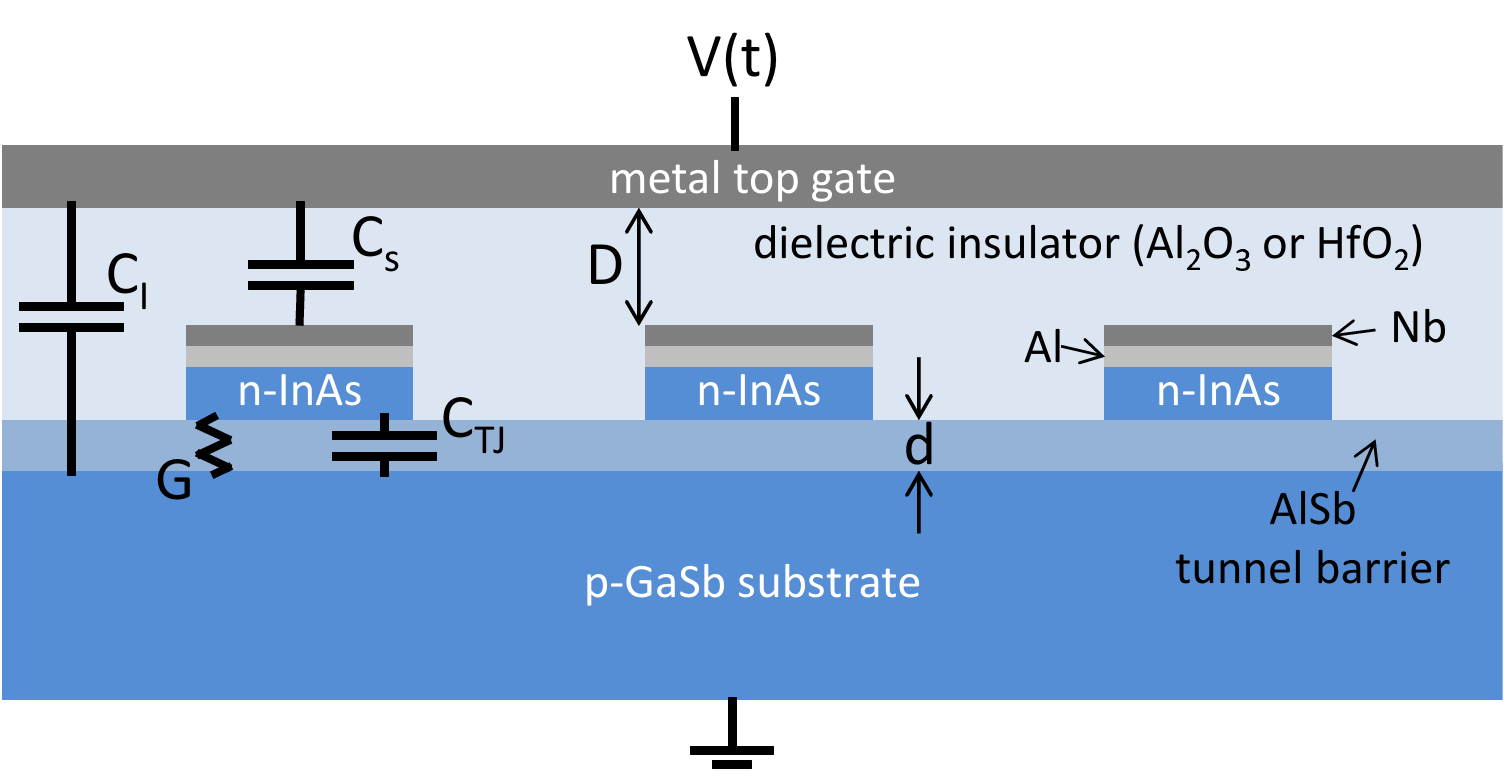}
\caption{Side-view of the proposed hybrid device geometry, showing cross-section of nanowires capacitively-coupled from above, with tunnel conductance pathway to the substrate below. The equivalent circuit diagram is overlayed on the left.
\label{fig1}}
\end{figure}

The conductance of a tunnel junction (TJ) between a normal bulk 3-dimensional metal and a superconductor reflects the superconducting gap in density of states near the Fermi energy, $\Delta$.\cite{Giaever_PRL1960} In our case, the superconductor is a one-dimensional proximity-coupled nanowire and cannot be ohmically contacted with a DC probe. It can, however, be \emph{capacitively} coupled to a counter electrode. This device, illustrated in side-view in Fig. 1, then has an equivalent circuit diagram where the tunnel contact to the nanowire consists of both a dissipative ($G$) and capacitive ($C_{TJ}$) path in parallel. We imagine an epitaxially-grown, lithographically-defined array of narrow-gap nanowires fabricated from a planar heterostructure (such as Al/InAs/AlSb/GaSb as shown in the figure), patterned with Nb and selectively etched, encapsulated in conformally-deposited dielectric insulator (such as atomic-layer-deposited Al$_2$O$_3$), and capped with a planar ``top gate'' metal counterelectrode capacitively coupled to the layers beneath it.\footnote{At sufficiently high DC bias, it may be possible to alter the chemical potential of the wire. This will be especially important to tune $\mu\approx 0$ to achieve the topological condition at magnetic fields low enough to preserve the superconducting state in the adjacent metal.} We further note that giant bandgap bowing in the InAsSb system\cite{Sarney_JVST2012, Belenky_APL2011, Belenky_APL2013,Wang_ProcSPIE2012} indicates that III-V materials with even larger spin-orbit interaction and smaller effective mass (both of which are essential in ensuring a wide topological gap) than InSb may be available for investigation. 

To illustrate the basic concept of the measurement scheme analytically, consider the \emph{nonlinear} conductor and a capacitor $C_{TJ}$ in parallel, shown in Fig. \ref{fig1}. When this system is used as a circuit-element model of an ideal tunnel junction, only odd powers of voltage are present in the current-voltage relation by symmetry: To leading order, the conductor current response to a voltage $V$ is described by the characteristic expression $G_1V+G_3V^3$, where $G_{1,3}$ are the linear and nonlinear conductances (with appropriately different units).

Ignoring the direct electrode-electrode capacitance $C_I$, we can equate the current flow in our hybrid device: 

\begin{equation}
G_1V_W+G_3V_W^3+i\omega C_{TJ}V_W=i\omega C_SV_C,\nonumber
\end{equation}

\noindent where $V_W$ is the voltage on the wire across the tunnel junction from the grounded bottom electrode and $V_C$ is the voltage across the purely capacitive component $C_S$ due to coupling to the wire from the top contact. By summing voltage, we also have
 
\begin{equation}
V_W+V_C=V(t)=V_W\left(1+\frac{G_1+G_3V_W^2+i\omega C_{TJ}}{i\omega C_S}\right).\nonumber
\end{equation}

If our driving voltage is $V(t)=\tilde{V}e^{i\omega t}$, then the system will find steady state with voltage oscillation across the tunnel junction of $V_W(t)=V_W^{(1)}e^{i\omega t}+V_W^{(3)}e^{i3\omega t}$; the response at the driving frequency $\omega$ is

\begin{equation}
V_W^{(1)}=\tilde{V}\frac{i\omega C_S}{G_1+i\omega (C_S+C_{TJ})}\nonumber
\end{equation}

\noindent and the \emph{third harmonic} amplitude is given by

\begin{equation}
V_W^{(3)}=(V_W^{(1)})^3\frac{-G_3}{G_1+i\omega(C_S+C_{TJ})}.\nonumber
\end{equation}

The current flowing through the circuit then also has a third harmonic component, 
\begin{equation}
i\omega C_SV_C^{(3)}=(V_W^{(1)})^3\frac{i\omega C_SG_3}{G_1+i\omega(C_S+C_{TJ})}.\nonumber
\end{equation}

\noindent Note that this response is linearly proportional to the nonlinear component of conductance, $G_3$. By measuring the third harmonic amplitude $V_W^{(3)}$ at $3\omega$ as a function of driving amplitude $| \tilde{V}|$ at $1\omega$, the higher-order nonlinear components present near the superconducting gap of the proximity-coupled nanowire can therefore be determined, \emph{despite our inability to maintain a DC bias across the tunnel junction} in Fig. \ref{fig1}. 

To model the relevant experimental aspects of this harmonic generation scheme, we have performed an explicit time-domain simulation of the high-frequency response of a simplified one-dimensional system incorporating the discrete spectrum of a finite wire. The geometry includes capacitive coupling from both the planar electrodes to the wire, and tunnel coupling to the grounded electrode, but ignores the direct electrode-electrode coupling. This simplification enables the implementation of a one-dimensional finite-differences Poisson equation to algebraically determine the electrostatic potential $V_W(t)$ on the floating wire:

\begin{equation}
-\nabla^2 V_W(t)=-\frac{\frac{V(t)-V_W(t)}{D}-\frac{V(t)}{d}}{(d+D)/2}=\frac{\rho}{\epsilon}=\frac{-2qN(t)}{\epsilon (d+D)\cdot A}, \nonumber
\end{equation}

\noindent where $D$ is the distance from the wire to the capacitive electrode at potential $V(t)=\tilde{V}sin(\omega t)$, $d$ is the distance to the tunnel electrode at ground, $N(t)$ is the number of electrons charging the wire out of equilibrium, $\epsilon$ is the dielectric permittivity between the wire and electrodes, $q$ is the fundamental charge, and $A$ is the cross-sectional area (wire length $\times$ width). This expression yields
 
\begin{equation}
V_W(t)=\left(\frac{V(t)}{D}-\frac{qN(t)}{\epsilon\cdot A}\right) \frac{dD}{d+D}.
\label{algebraicVeqn}
\end{equation}

After this wire voltage is calculated, we can determine the instantaneous current flowing into the wire $I=V\cdot G(V)$, where $G(V)$ is the voltage-dependent conductance to the wire. In a timestep $\Delta t$, this current imparts a transfer of charge $q\Delta N=I\cdot dt$, requiring repeated calculation of wire voltage, then current flow, etc. at every timestep. 

From this wire voltage fluctuation in time, one can then determine the displacement current flowing between the capacitive electrode (at voltage $V$) and the wire (at voltage $V_W$) by using the Maxwell addition to Ampere's law

\begin{equation}
I_d=A\cdot \epsilon \frac{\partial \mathcal{E}}{\partial t}=A\cdot \epsilon \frac{\partial}{\partial t}\left( \frac{V_W(t)-V(t)}{D} \right).
\label{displacementcurrenteqn} 
\end{equation}

\noindent Comparison of the magnitude and phase of this signal to the sinusoidal driving voltage $V(t)$ can be used to determine e.g. the complex impedance at a given simulation frequency $\omega$. The scheme has been benchmarked by considering a wire with continuous energy spectrum and linear current-voltage relationship with the tunnel contact, in which case the equivalent predictions of simple circuit theory are asymptotically obtained in the limit $\Delta t\rightarrow 0$.

\begin{figure}
\centering
\includegraphics[width=3.25in]{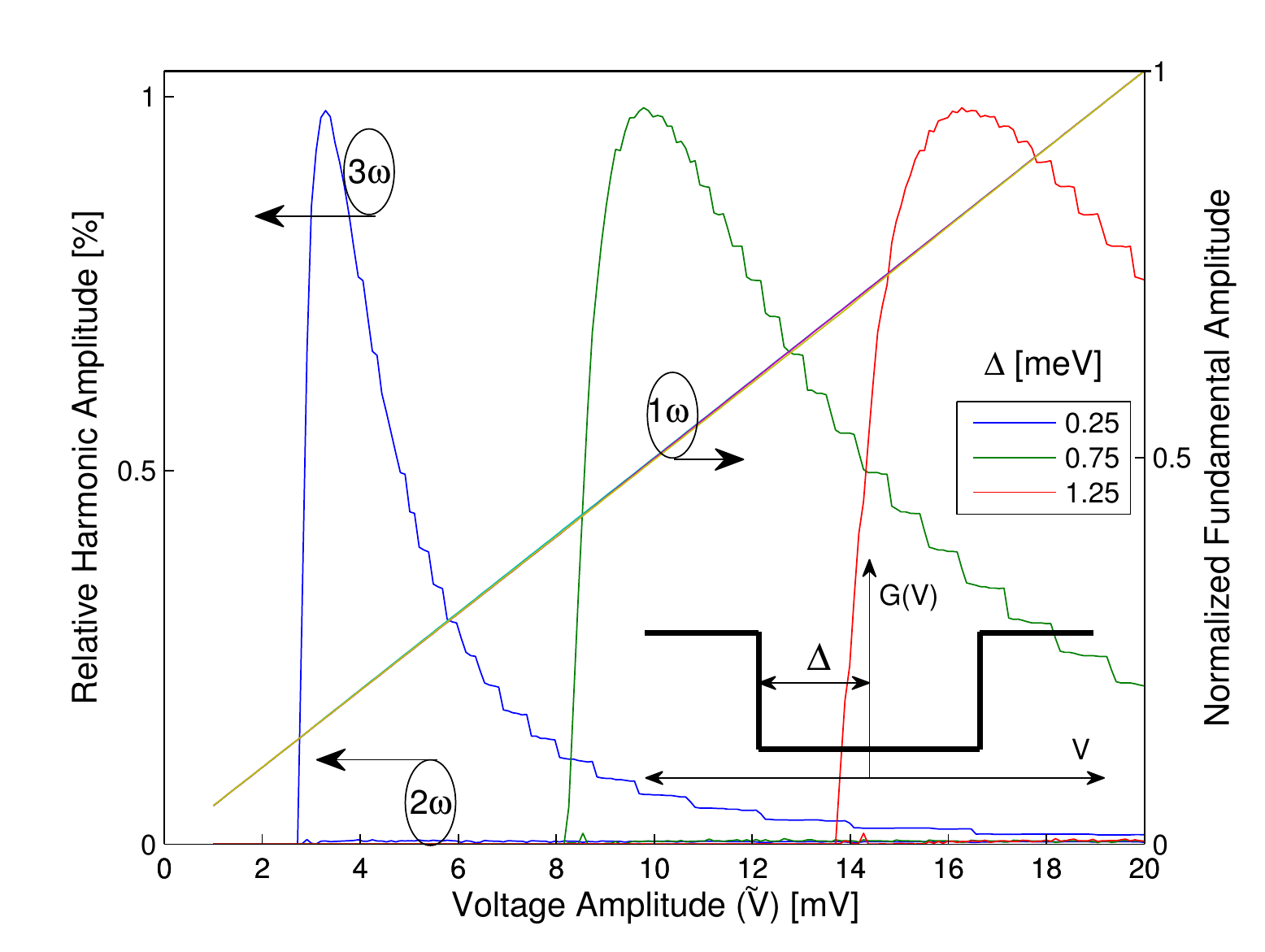}
\caption{Harmonic generation in time-domain simulation of the hybrid capacitor with nonlinear conductance spectrum at 5~MHz as shown in the inset. Whereas the fundamental ``1$\omega$" and second harmonic ``2$\omega$'' signals are relatively unaffected, the 3$\omega$ response reflects the gap energy $\Delta$, establishing the utility of this method to measure proximity gap closure of a nanowire in a magnetic field. 
\label{Figdelta}}
\end{figure}

We use the fast Fourier transform to determine the magnitude of third-harmonic $3\omega$ response of a one-dimensional wire. As discussed previously, we expect this signal as a function of oscillation amplitude $\tilde{V}$ at fundamental frequency $\omega$ to reflect the bias dependence of the nonlinear tunnel conductance spectrum.

To test this approach, we simulate the response of the hybrid capacitor with a simple nonlinear gapped conductance similar to what is expected from tunneling into a trivial superconductor.  Here, the device geometry has $d=$10~nm, $D$=100~nm, permittivity $\epsilon=12\epsilon_0$, and area= width$\times$length = 100nm $\times$ 1$\mu$m. The tunnel resistance varies from 1G$\Omega$ below the gap voltage $|qV|<\Delta$ to 0.1G$\Omega$ above it. As shown in Fig. \ref{Figdelta}, no signature of underlying spectrum features can be seen in the fundamental ``$1\omega$'' response at 5~MHz, and $2\omega$ signals are absent as expected by symmetry, even for multiple gap values $\Delta$. However, the $3\omega$ amplitude is approximately 1\% of the fundamental, and exhibits a sharp threshold at a voltage amplitude which scales linearly with $\Delta$. We are therefore confident in interpreting the 3$\omega$ signal as a function of excitation amplitude $\tilde{V}$ at frequency $\omega$ to be proportional to the nonlinear components of tunnel conductance spectrum in this hybrid tunnel/capacitor geometry.

\begin{figure}
\centering
\includegraphics[width=3.5in]{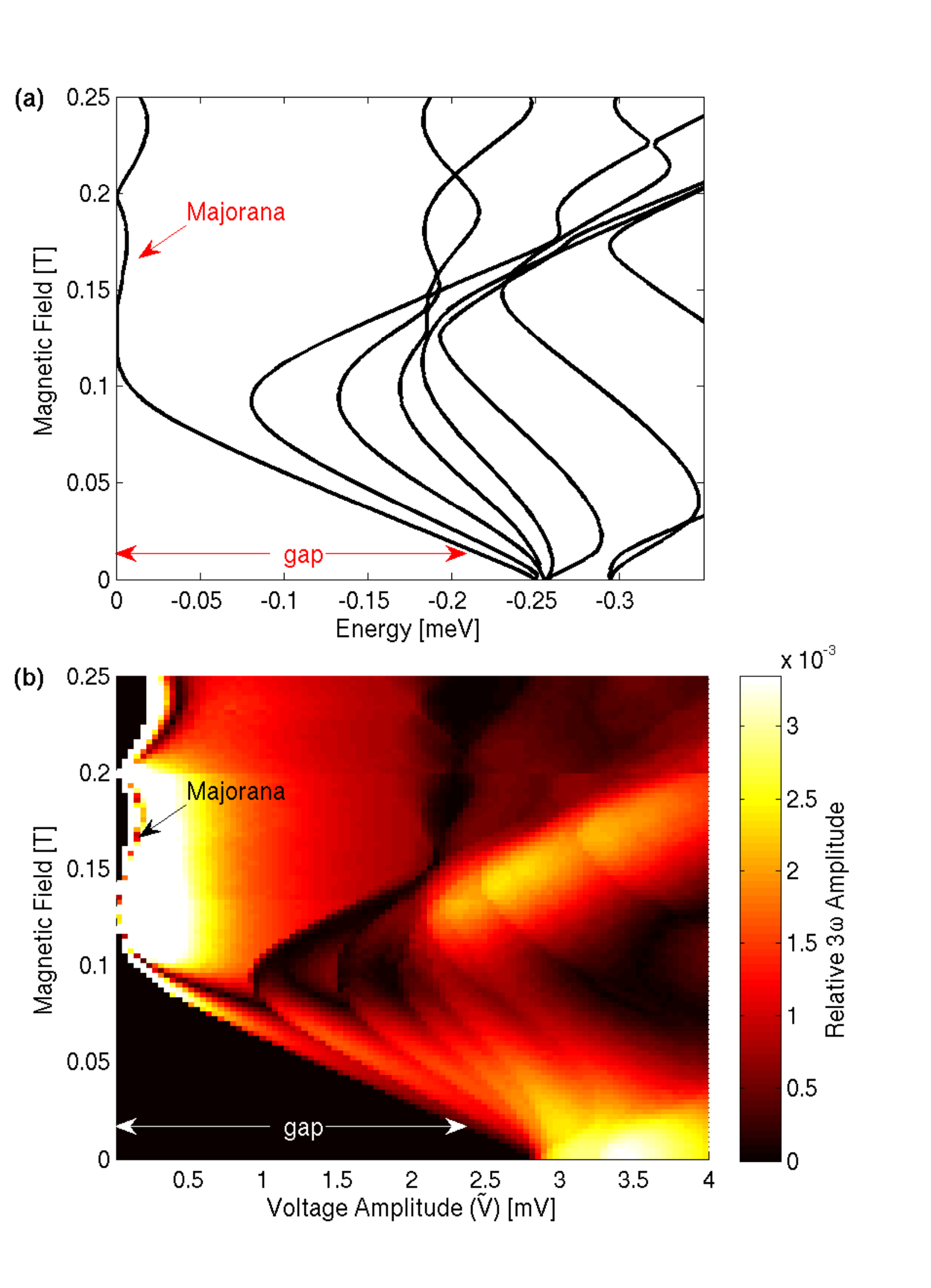}
\caption{(a) Energy spectrum of the filled states of a spin-orbit-coupled nanowire relative to the Fermi energy as a function of magnetic field, calculated by diagonalization of the finite-differences Bogoliubov-de~Gennes hamiltonian. (b)  Calculated amplitude of the third harmonic at $3\omega$ in the time-dependent response from a capacitively coupled nanowire with this excitation spectrum, relative to fundamental at $\omega$. 
\label{FigWire}}
\end{figure}

The incorporation of a discrete nanowire spectrum into the time-domain simulation involves a modification of the tunnel current expression, which can be calculated easily at zero temperature: 

\begin{equation}
I=e\frac{\gamma}{\hbar}\sum_i[\Theta(eV-E_i)-\Theta(-E_i)]\int|\Psi_i^e|^2dx,
\label{discreteIeqn}
\end{equation}

\noindent where $\gamma$ is coupling energy from the tunnel electrode, $\Theta$ is the Heaviside step function, and the sum counts all the $E_i$ discrete energy states of the wire that are available for transport due to occupation and Pauli exclusion. $|\Psi_i^e|^2$ is the electron-like probability of the wavefunction with eigenenergy $E_i$, relevant for hamiltonians which include superconductivity pairing between electrons and holes. The excitation spectrum has been calculated by numerical diagonalization of the single-band 1-dimensional Bogoliubov-de~Gennes hamiltonian incorporating Zeeman effect and spin-orbit interaction\cite{Sau_PRL2010, Sau_PRB2010, Oreg_PRL2010, Rainis_PRB2013} for a 1-micron-long wire with 40 spatial discretizations (yielding a 160$\times$160 matrix encoded using sparse matrix manipulation in Matlab) with the following parameters: spin-orbit coefficient $\alpha=$0.2~eV$\AA$, chemical potential $\mu=$0 meV (at the spin-degeneracy Dirac point in the absence of magnetic field), superconductivity pairing energy $\Delta=$0.25~meV, effective mass $m^*=0.026m_0$, and g-factor $g=50$.

In Figure \ref{FigWire}(a), we show the energy spectrum evolution with magnetic field $B$. When the topological condition for a helical state $g\mu_B B > \sqrt{\mu^2 + \Delta ^2}$ is achieved, the superconducting gap disappears due to the coalescence of energies driven by Zeeman effect; beyond this magnetic field value, nearly degenerate states at the Fermi energy are seen. Due to the finite wire length, interactions between the end-localized Majorana states cause level repulsion and oscillation - the so-called ``smoking gun'' of the Majorana.\cite{DasSarma_PRB2012} 

Figure \ref{FigWire}(b) shows the resulting $3\omega$ response of a hybrid capacitor using this spectrum and Eqs. \ref{algebraicVeqn}, \ref{displacementcurrenteqn}, and \ref{discreteIeqn} at 5~MHz. Again, the device geometry has $d=$10~nm, $D$=100~nm, permittivity $\epsilon=12\epsilon_0$, and area= width$\times$length = 100nm $\times$ 1$\mu$m. The tunnel coupling constant $\gamma$ is 1neV, much smaller than the gap $\Delta$, justifying our disregard for state broadening; this coupling value is equivalent to a tunnel resistance $R\approx\frac{\hbar}{e^2}\frac{\Delta}{\gamma}\approx $1~G$\Omega$, which is necessarily large to constrain the flow of current in a single timestep of $\Delta t=$12.6~ns $<$ 1 electron.

All the spectrum features are captured directly in this $3\omega$ harmonic generation. The gap can be seen to close and a strong signal at low excitation amplitudes corresponds to the creation of the Majorana at the Fermi energy. The strength of the third harmonic under conditions corresponding to the continuum/gap edge is approximately $0.3\%$ of the fundamental excitation amplitude. We expect this robust signal will be experimentally measurable with an appropriate lock-in amplifier or spectrum analyzer. 

The experimental feasibility of the approach to measure tunnel conductance spectra outlined here can be established in a planar device with an equivalent electrical configuration, but with a trivial superconducting gap. For example, fabrication of a Cu/AlO$_x$/Al tunnel junction on an oxidized Si substrate with a thin insulating layer of SiO$_2$ will provide exactly the same configuration shown in Fig. 1, but upside-down: The capacitive counter electrode is now the substrate, and the tunnel contact is the top thin-film layer. 

This control experiment will be especially useful in refining the scheme, as direct contact to the planar superconductor (not possible for the Majorana wire) can be used to independently calibrate using conventional conductance spectroscopy. At low temperatures, the energy gap observed through harmonic generation should close in modest magnetic fields, and no signature of a Majorana ZBC should be seen. 

We end on a caveat: In an imperfect tunnel junction, current contributions of the form $G_2V^2$ (Ref. \onlinecite{Brinkman_JAP1970}) will pollute the 3$\omega$ signal via terms proportional to $G_1\times G_2$. Care must therefore be taken in the materials growth to assure a symmetric barrier potential so that $G_2=0$.  
  
\begin{acknowledgments}
The author acknowledges important conversations with T. Stanescu, J.D. Sau, B. Halperin, and G. Ben-Shach, and the generous hospitality of Prof. A. Yacoby at Harvard during the preparation of this manuscript. 
\end{acknowledgments}


\begin{thebibliography}{27}%
\makeatletter
\providecommand \@ifxundefined [1]{%
 \@ifx{#1\undefined}
}%
\providecommand \@ifnum [1]{%
 \ifnum #1\expandafter \@firstoftwo
 \else \expandafter \@secondoftwo
 \fi
}%
\providecommand \@ifx [1]{%
 \ifx #1\expandafter \@firstoftwo
 \else \expandafter \@secondoftwo
 \fi
}%
\providecommand \natexlab [1]{#1}%
\providecommand \enquote  [1]{``#1''}%
\providecommand \bibnamefont  [1]{#1}%
\providecommand \bibfnamefont [1]{#1}%
\providecommand \citenamefont [1]{#1}%
\providecommand \href@noop [0]{\@secondoftwo}%
\providecommand \href [0]{\begingroup \@sanitize@url \@href}%
\providecommand \@href[1]{\@@startlink{#1}\@@href}%
\providecommand \@@href[1]{\endgroup#1\@@endlink}%
\providecommand \@sanitize@url [0]{\catcode `\\12\catcode `\$12\catcode
  `\&12\catcode `\#12\catcode `\^12\catcode `\_12\catcode `\%12\relax}%
\providecommand \@@startlink[1]{}%
\providecommand \@@endlink[0]{}%
\providecommand \url  [0]{\begingroup\@sanitize@url \@url }%
\providecommand \@url [1]{\endgroup\@href {#1}{\urlprefix }}%
\providecommand \urlprefix  [0]{URL }%
\providecommand \Eprint [0]{\href }%
\providecommand \doibase [0]{http://dx.doi.org/}%
\providecommand \selectlanguage [0]{\@gobble}%
\providecommand \bibinfo  [0]{\@secondoftwo}%
\providecommand \bibfield  [0]{\@secondoftwo}%
\providecommand \translation [1]{[#1]}%
\providecommand \BibitemOpen [0]{}%
\providecommand \bibitemStop [0]{}%
\providecommand \bibitemNoStop [0]{.\EOS\space}%
\providecommand \EOS [0]{\spacefactor3000\relax}%
\providecommand \BibitemShut  [1]{\csname bibitem#1\endcsname}%
\let\auto@bib@innerbib\@empty
\bibitem [{\citenamefont {Mourik}\ \emph {et~al.}(2012)\citenamefont {Mourik},
  \citenamefont {Zuo}, \citenamefont {Frolov}, \citenamefont {Plissard},
  \citenamefont {Bakkers},\ and\ \citenamefont
  {Kouwenhoven}}]{Mourik_Science2012}%
  \BibitemOpen
  \bibfield  {author} {\bibinfo {author} {\bibfnamefont {V.}~\bibnamefont
  {Mourik}}, \bibinfo {author} {\bibfnamefont {K.}~\bibnamefont {Zuo}},
  \bibinfo {author} {\bibfnamefont {S.~M.}\ \bibnamefont {Frolov}}, \bibinfo
  {author} {\bibfnamefont {S.}~\bibnamefont {Plissard}}, \bibinfo {author}
  {\bibfnamefont {E.~A.}\ \bibnamefont {Bakkers}}, \ and\ \bibinfo {author}
  {\bibfnamefont {L.}~\bibnamefont {Kouwenhoven}},\ }\href@noop {} {\bibfield
  {journal} {\bibinfo  {journal} {Science}\ }\textbf {\bibinfo {volume}
  {336}},\ \bibinfo {pages} {1003} (\bibinfo {year} {2012})}\BibitemShut
  {NoStop}%
\bibitem [{\citenamefont {Das}\ \emph {et~al.}(2012)\citenamefont {Das},
  \citenamefont {Ronen}, \citenamefont {Most}, \citenamefont {Oreg},
  \citenamefont {Heiblum},\ and\ \citenamefont
  {Shtrikman}}]{Das_NaturePhys2012}%
  \BibitemOpen
  \bibfield  {author} {\bibinfo {author} {\bibfnamefont {A.}~\bibnamefont
  {Das}}, \bibinfo {author} {\bibfnamefont {Y.}~\bibnamefont {Ronen}}, \bibinfo
  {author} {\bibfnamefont {Y.}~\bibnamefont {Most}}, \bibinfo {author}
  {\bibfnamefont {Y.}~\bibnamefont {Oreg}}, \bibinfo {author} {\bibfnamefont
  {M.}~\bibnamefont {Heiblum}}, \ and\ \bibinfo {author} {\bibfnamefont
  {H.}~\bibnamefont {Shtrikman}},\ }\href@noop {} {\bibfield  {journal}
  {\bibinfo  {journal} {Nature Phys.}\ }\textbf {\bibinfo {volume} {8}},\
  \bibinfo {pages} {887} (\bibinfo {year} {2012})}\BibitemShut {NoStop}%
\bibitem [{\citenamefont {Deng}\ \emph {et~al.}(2012)\citenamefont {Deng},
  \citenamefont {Yu}, \citenamefont {Huang}, \citenamefont {Larsson},
  \citenamefont {Caroff},\ and\ \citenamefont {Xu}}]{Deng_NanoLett2012}%
  \BibitemOpen
  \bibfield  {author} {\bibinfo {author} {\bibfnamefont {M.~T.}\ \bibnamefont
  {Deng}}, \bibinfo {author} {\bibfnamefont {C.~L.}\ \bibnamefont {Yu}},
  \bibinfo {author} {\bibfnamefont {G.~Y.}\ \bibnamefont {Huang}}, \bibinfo
  {author} {\bibfnamefont {M.}~\bibnamefont {Larsson}}, \bibinfo {author}
  {\bibfnamefont {P.}~\bibnamefont {Caroff}}, \ and\ \bibinfo {author}
  {\bibfnamefont {H.~Q.}\ \bibnamefont {Xu}},\ }\href@noop {} {\bibfield
  {journal} {\bibinfo  {journal} {Nano Lett.}\ }\textbf {\bibinfo {volume}
  {12}},\ \bibinfo {pages} {6414} (\bibinfo {year} {2012})}\BibitemShut
  {NoStop}%
\bibitem [{\citenamefont {Nayak}\ \emph {et~al.}(2008)\citenamefont {Nayak},
  \citenamefont {Simon}, \citenamefont {Stern}, \citenamefont {Freedman},\ and\
  \citenamefont {Sarma}}]{Nayak_RMP2008}%
  \BibitemOpen
  \bibfield  {author} {\bibinfo {author} {\bibfnamefont {C.}~\bibnamefont
  {Nayak}}, \bibinfo {author} {\bibfnamefont {S.~H.}\ \bibnamefont {Simon}},
  \bibinfo {author} {\bibfnamefont {A.}~\bibnamefont {Stern}}, \bibinfo
  {author} {\bibfnamefont {M.}~\bibnamefont {Freedman}}, \ and\ \bibinfo
  {author} {\bibfnamefont {S.~D.}\ \bibnamefont {Sarma}},\ }\href@noop {}
  {\bibfield  {journal} {\bibinfo  {journal} {Rev. Mod. Phys.}\ }\textbf
  {\bibinfo {volume} {80}},\ \bibinfo {pages} {1083} (\bibinfo {year}
  {2008})}\BibitemShut {NoStop}%
\bibitem [{\citenamefont {Kitaev}(2001)}]{Kitaev_Uspekhi2001}%
  \BibitemOpen
  \bibfield  {author} {\bibinfo {author} {\bibfnamefont {A.~Y.}\ \bibnamefont
  {Kitaev}},\ }\href@noop {} {\bibfield  {journal} {\bibinfo  {journal}
  {Physics-Uspekhi}\ }\textbf {\bibinfo {volume} {44}},\ \bibinfo {pages} {131}
  (\bibinfo {year} {2001})}\BibitemShut {NoStop}%
\bibitem [{\citenamefont {Read}\ and\ \citenamefont
  {Green}(2000)}]{Read_PRB2000}%
  \BibitemOpen
  \bibfield  {author} {\bibinfo {author} {\bibfnamefont {N.}~\bibnamefont
  {Read}}\ and\ \bibinfo {author} {\bibfnamefont {D.}~\bibnamefont {Green}},\
  }\href {http://link.aps.org/doi/10.1103/PhysRevB.61.10267} {\bibfield
  {journal} {\bibinfo  {journal} {Phys. Rev. B}\ }\textbf {\bibinfo {volume}
  {61}},\ \bibinfo {pages} {10267} (\bibinfo {year} {2000})}\BibitemShut
  {NoStop}%
\bibitem [{\citenamefont {Sau}\ \emph {et~al.}(2010{\natexlab{a}})\citenamefont
  {Sau}, \citenamefont {Lutchyn}, \citenamefont {Tewari},\ and\ \citenamefont
  {Das~Sarma}}]{Sau_PRL2010}%
  \BibitemOpen
  \bibfield  {author} {\bibinfo {author} {\bibfnamefont {J.~D.}\ \bibnamefont
  {Sau}}, \bibinfo {author} {\bibfnamefont {R.~M.}\ \bibnamefont {Lutchyn}},
  \bibinfo {author} {\bibfnamefont {S.}~\bibnamefont {Tewari}}, \ and\ \bibinfo
  {author} {\bibfnamefont {S.}~\bibnamefont {Das~Sarma}},\ }\href
  {http://link.aps.org/doi/10.1103/PhysRevLett.104.040502} {\bibfield
  {journal} {\bibinfo  {journal} {Phys. Rev. Lett.}\ }\textbf {\bibinfo
  {volume} {104}},\ \bibinfo {pages} {040502} (\bibinfo {year}
  {2010}{\natexlab{a}})}\BibitemShut {NoStop}%
\bibitem [{\citenamefont {Sengupta}\ \emph {et~al.}(2001)\citenamefont
  {Sengupta}, \citenamefont {\ifmmode \check{Z}\else
  \v{Z}\fi{}uti\ifmmode~\acute{c}\else \'{c}\fi{}}, \citenamefont {Kwon},
  \citenamefont {Yakovenko},\ and\ \citenamefont
  {Das~Sarma}}]{Sengupta_PRB2001}%
  \BibitemOpen
  \bibfield  {author} {\bibinfo {author} {\bibfnamefont {K.}~\bibnamefont
  {Sengupta}}, \bibinfo {author} {\bibfnamefont {I.}~\bibnamefont {\ifmmode
  \check{Z}\else \v{Z}\fi{}uti\ifmmode~\acute{c}\else \'{c}\fi{}}}, \bibinfo
  {author} {\bibfnamefont {H.-J.}\ \bibnamefont {Kwon}}, \bibinfo {author}
  {\bibfnamefont {V.~M.}\ \bibnamefont {Yakovenko}}, \ and\ \bibinfo {author}
  {\bibfnamefont {S.}~\bibnamefont {Das~Sarma}},\ }\href
  {http://link.aps.org/doi/10.1103/PhysRevB.63.144531} {\bibfield  {journal}
  {\bibinfo  {journal} {Phys. Rev. B}\ }\textbf {\bibinfo {volume} {63}},\
  \bibinfo {pages} {144531} (\bibinfo {year} {2001})}\BibitemShut {NoStop}%
\bibitem [{\citenamefont {Flensberg}(2010)}]{Flensberg_PRB2010}%
  \BibitemOpen
  \bibfield  {author} {\bibinfo {author} {\bibfnamefont {K.}~\bibnamefont
  {Flensberg}},\ }\href {http://link.aps.org/doi/10.1103/PhysRevB.82.180516}
  {\bibfield  {journal} {\bibinfo  {journal} {Phys. Rev. B}\ }\textbf {\bibinfo
  {volume} {82}},\ \bibinfo {pages} {180516} (\bibinfo {year}
  {2010})}\BibitemShut {NoStop}%
\bibitem [{\citenamefont {Alicea}(2010)}]{Alicea_PRB2010}%
  \BibitemOpen
  \bibfield  {author} {\bibinfo {author} {\bibfnamefont {J.}~\bibnamefont
  {Alicea}},\ }\href {http://link.aps.org/doi/10.1103/PhysRevB.81.125318}
  {\bibfield  {journal} {\bibinfo  {journal} {Phys. Rev. B}\ }\textbf {\bibinfo
  {volume} {81}},\ \bibinfo {pages} {125318} (\bibinfo {year}
  {2010})}\BibitemShut {NoStop}%
\bibitem [{\citenamefont {Wimmer}\ \emph {et~al.}(2011)\citenamefont {Wimmer},
  \citenamefont {Akhmerov}, \citenamefont {Dahlhaus},\ and\ \citenamefont
  {Beenakker}}]{Wimmer_NJP2011}%
  \BibitemOpen
  \bibfield  {author} {\bibinfo {author} {\bibfnamefont {M.}~\bibnamefont
  {Wimmer}}, \bibinfo {author} {\bibfnamefont {A.}~\bibnamefont {Akhmerov}},
  \bibinfo {author} {\bibfnamefont {J.}~\bibnamefont {Dahlhaus}}, \ and\
  \bibinfo {author} {\bibfnamefont {C.}~\bibnamefont {Beenakker}},\ }\href@noop
  {} {\bibfield  {journal} {\bibinfo  {journal} {New J. Phys.}\ }\textbf
  {\bibinfo {volume} {13}},\ \bibinfo {pages} {053016} (\bibinfo {year}
  {2011})}\BibitemShut {NoStop}%
\bibitem [{\citenamefont {Stanescu}\ \emph {et~al.}(2011)\citenamefont
  {Stanescu}, \citenamefont {Lutchyn},\ and\ \citenamefont
  {Das~Sarma}}]{Stanescu_PRB2011}%
  \BibitemOpen
  \bibfield  {author} {\bibinfo {author} {\bibfnamefont {T.~D.}\ \bibnamefont
  {Stanescu}}, \bibinfo {author} {\bibfnamefont {R.~M.}\ \bibnamefont
  {Lutchyn}}, \ and\ \bibinfo {author} {\bibfnamefont {S.}~\bibnamefont
  {Das~Sarma}},\ }\href {http://link.aps.org/doi/10.1103/PhysRevB.84.144522}
  {\bibfield  {journal} {\bibinfo  {journal} {Phys. Rev. B}\ }\textbf {\bibinfo
  {volume} {84}},\ \bibinfo {pages} {144522} (\bibinfo {year}
  {2011})}\BibitemShut {NoStop}%
\bibitem [{\citenamefont {Law}\ \emph {et~al.}(2009)\citenamefont {Law},
  \citenamefont {Lee},\ and\ \citenamefont {Ng}}]{Law_PRL2009}%
  \BibitemOpen
  \bibfield  {author} {\bibinfo {author} {\bibfnamefont {K.~T.}\ \bibnamefont
  {Law}}, \bibinfo {author} {\bibfnamefont {P.~A.}\ \bibnamefont {Lee}}, \ and\
  \bibinfo {author} {\bibfnamefont {T.~K.}\ \bibnamefont {Ng}},\ }\href
  {http://link.aps.org/doi/10.1103/PhysRevLett.103.237001} {\bibfield
  {journal} {\bibinfo  {journal} {Phys. Rev. Lett.}\ }\textbf {\bibinfo
  {volume} {103}},\ \bibinfo {pages} {237001} (\bibinfo {year}
  {2009})}\BibitemShut {NoStop}%
\bibitem [{\citenamefont {Stanescu}\ \emph {et~al.}(2012)\citenamefont
  {Stanescu}, \citenamefont {Tewari}, \citenamefont {Sau},\ and\ \citenamefont
  {Das~Sarma}}]{Stanescu_PRL2012}%
  \BibitemOpen
  \bibfield  {author} {\bibinfo {author} {\bibfnamefont {T.~D.}\ \bibnamefont
  {Stanescu}}, \bibinfo {author} {\bibfnamefont {S.}~\bibnamefont {Tewari}},
  \bibinfo {author} {\bibfnamefont {J.~D.}\ \bibnamefont {Sau}}, \ and\
  \bibinfo {author} {\bibfnamefont {S.}~\bibnamefont {Das~Sarma}},\ }\href
  {http://link.aps.org/doi/10.1103/PhysRevLett.109.266402} {\bibfield
  {journal} {\bibinfo  {journal} {Phys. Rev. Lett.}\ }\textbf {\bibinfo
  {volume} {109}},\ \bibinfo {pages} {266402} (\bibinfo {year}
  {2012})}\BibitemShut {NoStop}%
\bibitem [{\citenamefont {Shapiro}(1967)}]{Shapiro_JAP1967}%
  \BibitemOpen
  \bibfield  {author} {\bibinfo {author} {\bibfnamefont {S.}~\bibnamefont
  {Shapiro}},\ }\href {\doibase 10.1063/1.1709777} {\bibfield  {journal}
  {\bibinfo  {journal} {J. Appl. Phys.}\ }\textbf {\bibinfo {volume} {38}},\
  \bibinfo {pages} {1879} (\bibinfo {year} {1967})}\BibitemShut {NoStop}%
\bibitem [{\citenamefont {Reuther}\ \emph {et~al.}(2013)\citenamefont
  {Reuther}, \citenamefont {Alicea},\ and\ \citenamefont
  {Yacoby}}]{Reuther_arxiv2013}%
  \BibitemOpen
  \bibfield  {author} {\bibinfo {author} {\bibfnamefont {J.}~\bibnamefont
  {Reuther}}, \bibinfo {author} {\bibfnamefont {J.}~\bibnamefont {Alicea}}, \
  and\ \bibinfo {author} {\bibfnamefont {A.}~\bibnamefont {Yacoby}},\
  }\href@noop {} {\enquote {\bibinfo {title} {Gate defined wires in hgte
  quantum wells: from majorana fermions to spintronics},}\ } (\bibinfo {year}
  {2013}),\ \Eprint {http://arxiv.org/abs/cond-mat/1303.1207}
  {arXiv:cond-mat/1303.1207} \BibitemShut {NoStop}%
\bibitem [{\citenamefont {Giaever}(1960)}]{Giaever_PRL1960}%
  \BibitemOpen
  \bibfield  {author} {\bibinfo {author} {\bibfnamefont {I.}~\bibnamefont
  {Giaever}},\ }\href {http://link.aps.org/doi/10.1103/PhysRevLett.5.147}
  {\bibfield  {journal} {\bibinfo  {journal} {Phys. Rev. Lett.}\ }\textbf
  {\bibinfo {volume} {5}},\ \bibinfo {pages} {147} (\bibinfo {year}
  {1960})}\BibitemShut {NoStop}%
\bibitem [{Note1()}]{Note1}%
  \BibitemOpen
  \bibinfo {note} {At sufficiently high DC bias, it may be possible to alter
  the chemical potential of the wire. This will be especially important to tune
  $\mu \approx 0$ to achieve the topological condition at magnetic fields low
  enough to preserve the superconducting state in the adjacent
  metal.}\BibitemShut {Stop}%
\bibitem [{\citenamefont {Sarney}\ \emph {et~al.}(2012)\citenamefont {Sarney},
  \citenamefont {Svensson}, \citenamefont {Hier}, \citenamefont {Kipshidze},
  \citenamefont {Donetsky}, \citenamefont {Wang}, \citenamefont {Shterengas},\
  and\ \citenamefont {Belenky}}]{Sarney_JVST2012}%
  \BibitemOpen
  \bibfield  {author} {\bibinfo {author} {\bibfnamefont {W.~L.}\ \bibnamefont
  {Sarney}}, \bibinfo {author} {\bibfnamefont {S.~P.}\ \bibnamefont
  {Svensson}}, \bibinfo {author} {\bibfnamefont {H.}~\bibnamefont {Hier}},
  \bibinfo {author} {\bibfnamefont {G.}~\bibnamefont {Kipshidze}}, \bibinfo
  {author} {\bibfnamefont {D.}~\bibnamefont {Donetsky}}, \bibinfo {author}
  {\bibfnamefont {D.}~\bibnamefont {Wang}}, \bibinfo {author} {\bibfnamefont
  {L.}~\bibnamefont {Shterengas}}, \ and\ \bibinfo {author} {\bibfnamefont
  {G.}~\bibnamefont {Belenky}},\ }\href
  {http://link.aip.org/link/?JVB/30/02B105/1} {\bibfield  {journal} {\bibinfo
  {journal} {J. Vac. Sci. Tech. B}\ }\textbf {\bibinfo {volume} {30}},\
  \bibinfo {pages} {02B105} (\bibinfo {year} {2012})}\BibitemShut {NoStop}%
\bibitem [{\citenamefont {Belenky}\ \emph {et~al.}(2011)\citenamefont
  {Belenky}, \citenamefont {Donetsky}, \citenamefont {Kipshidze}, \citenamefont
  {Wang}, \citenamefont {Shterengas}, \citenamefont {Sarney},\ and\
  \citenamefont {Svensson}}]{Belenky_APL2011}%
  \BibitemOpen
  \bibfield  {author} {\bibinfo {author} {\bibfnamefont {G.}~\bibnamefont
  {Belenky}}, \bibinfo {author} {\bibfnamefont {D.}~\bibnamefont {Donetsky}},
  \bibinfo {author} {\bibfnamefont {G.}~\bibnamefont {Kipshidze}}, \bibinfo
  {author} {\bibfnamefont {D.}~\bibnamefont {Wang}}, \bibinfo {author}
  {\bibfnamefont {L.}~\bibnamefont {Shterengas}}, \bibinfo {author}
  {\bibfnamefont {W.~L.}\ \bibnamefont {Sarney}}, \ and\ \bibinfo {author}
  {\bibfnamefont {S.~P.}\ \bibnamefont {Svensson}},\ }\href@noop {} {\bibfield
  {journal} {\bibinfo  {journal} {Appl. Phys. Lett.}\ }\textbf {\bibinfo
  {volume} {99}},\ \bibinfo {pages} {141116} (\bibinfo {year}
  {2011})}\BibitemShut {NoStop}%
\bibitem [{\citenamefont {Belenky}\ \emph {et~al.}(2013)\citenamefont
  {Belenky}, \citenamefont {Wang}, \citenamefont {Lin}, \citenamefont
  {Donetsky}, \citenamefont {Kipshidze}, \citenamefont {Shterengas},
  \citenamefont {Westerfeld}, \citenamefont {Sarney},\ and\ \citenamefont
  {Svensson}}]{Belenky_APL2013}%
  \BibitemOpen
  \bibfield  {author} {\bibinfo {author} {\bibfnamefont {G.}~\bibnamefont
  {Belenky}}, \bibinfo {author} {\bibfnamefont {D.}~\bibnamefont {Wang}},
  \bibinfo {author} {\bibfnamefont {Y.}~\bibnamefont {Lin}}, \bibinfo {author}
  {\bibfnamefont {D.}~\bibnamefont {Donetsky}}, \bibinfo {author}
  {\bibfnamefont {G.}~\bibnamefont {Kipshidze}}, \bibinfo {author}
  {\bibfnamefont {L.}~\bibnamefont {Shterengas}}, \bibinfo {author}
  {\bibfnamefont {D.}~\bibnamefont {Westerfeld}}, \bibinfo {author}
  {\bibfnamefont {W.}~\bibnamefont {Sarney}}, \ and\ \bibinfo {author}
  {\bibfnamefont {S.}~\bibnamefont {Svensson}},\ }\href@noop {} {\bibfield
  {journal} {\bibinfo  {journal} {Appl. Phys. Lett.}\ }\textbf {\bibinfo
  {volume} {102}},\ \bibinfo {pages} {111108} (\bibinfo {year}
  {2013})}\BibitemShut {NoStop}%
\bibitem [{\citenamefont {Wang}\ \emph {et~al.}(2012)\citenamefont {Wang},
  \citenamefont {Lin}, \citenamefont {Donetsky}, \citenamefont {Shterengas},
  \citenamefont {Kipshidze}, \citenamefont {Belenky}, \citenamefont {Sarney},
  \citenamefont {Hier},\ and\ \citenamefont {Svensson}}]{Wang_ProcSPIE2012}%
  \BibitemOpen
  \bibfield  {author} {\bibinfo {author} {\bibfnamefont {D.}~\bibnamefont
  {Wang}}, \bibinfo {author} {\bibfnamefont {Y.}~\bibnamefont {Lin}}, \bibinfo
  {author} {\bibfnamefont {D.}~\bibnamefont {Donetsky}}, \bibinfo {author}
  {\bibfnamefont {L.}~\bibnamefont {Shterengas}}, \bibinfo {author}
  {\bibfnamefont {G.}~\bibnamefont {Kipshidze}}, \bibinfo {author}
  {\bibfnamefont {G.}~\bibnamefont {Belenky}}, \bibinfo {author} {\bibfnamefont
  {W.~L.}\ \bibnamefont {Sarney}}, \bibinfo {author} {\bibfnamefont
  {H.}~\bibnamefont {Hier}}, \ and\ \bibinfo {author} {\bibfnamefont
  {S.}~\bibnamefont {Svensson}},\ }\href@noop {} {\bibfield  {journal}
  {\bibinfo  {journal} {Proc. SPIE}\ }\textbf {\bibinfo {volume} {8353}},\
  \bibinfo {pages} {835312} (\bibinfo {year} {2012})}\BibitemShut {NoStop}%
\bibitem [{\citenamefont {Sau}\ \emph {et~al.}(2010{\natexlab{b}})\citenamefont
  {Sau}, \citenamefont {Tewari}, \citenamefont {Lutchyn}, \citenamefont
  {Stanescu},\ and\ \citenamefont {Das~Sarma}}]{Sau_PRB2010}%
  \BibitemOpen
  \bibfield  {author} {\bibinfo {author} {\bibfnamefont {J.~D.}\ \bibnamefont
  {Sau}}, \bibinfo {author} {\bibfnamefont {S.}~\bibnamefont {Tewari}},
  \bibinfo {author} {\bibfnamefont {R.~M.}\ \bibnamefont {Lutchyn}}, \bibinfo
  {author} {\bibfnamefont {T.~D.}\ \bibnamefont {Stanescu}}, \ and\ \bibinfo
  {author} {\bibfnamefont {S.}~\bibnamefont {Das~Sarma}},\ }\href
  {http://link.aps.org/doi/10.1103/PhysRevB.82.214509} {\bibfield  {journal}
  {\bibinfo  {journal} {Phys. Rev. B}\ }\textbf {\bibinfo {volume} {82}},\
  \bibinfo {pages} {214509} (\bibinfo {year} {2010}{\natexlab{b}})}\BibitemShut
  {NoStop}%
\bibitem [{\citenamefont {Oreg}\ \emph {et~al.}(2010)\citenamefont {Oreg},
  \citenamefont {Refael},\ and\ \citenamefont {von Oppen}}]{Oreg_PRL2010}%
  \BibitemOpen
  \bibfield  {author} {\bibinfo {author} {\bibfnamefont {Y.}~\bibnamefont
  {Oreg}}, \bibinfo {author} {\bibfnamefont {G.}~\bibnamefont {Refael}}, \ and\
  \bibinfo {author} {\bibfnamefont {F.}~\bibnamefont {von Oppen}},\ }\href
  {http://link.aps.org/doi/10.1103/PhysRevLett.105.177002} {\bibfield
  {journal} {\bibinfo  {journal} {Phys. Rev. Lett.}\ }\textbf {\bibinfo
  {volume} {105}},\ \bibinfo {pages} {177002} (\bibinfo {year}
  {2010})}\BibitemShut {NoStop}%
\bibitem [{\citenamefont {Rainis}\ \emph {et~al.}(2013)\citenamefont {Rainis},
  \citenamefont {Trifunovic}, \citenamefont {Klinovaja},\ and\ \citenamefont
  {Loss}}]{Rainis_PRB2013}%
  \BibitemOpen
  \bibfield  {author} {\bibinfo {author} {\bibfnamefont {D.}~\bibnamefont
  {Rainis}}, \bibinfo {author} {\bibfnamefont {L.}~\bibnamefont {Trifunovic}},
  \bibinfo {author} {\bibfnamefont {J.}~\bibnamefont {Klinovaja}}, \ and\
  \bibinfo {author} {\bibfnamefont {D.}~\bibnamefont {Loss}},\ }\href
  {http://link.aps.org/doi/10.1103/PhysRevB.87.024515} {\bibfield  {journal}
  {\bibinfo  {journal} {Phys. Rev. B}\ }\textbf {\bibinfo {volume} {87}},\
  \bibinfo {pages} {024515} (\bibinfo {year} {2013})}\BibitemShut {NoStop}%
\bibitem [{\citenamefont {Das~Sarma}\ \emph {et~al.}(2012)\citenamefont
  {Das~Sarma}, \citenamefont {Sau},\ and\ \citenamefont
  {Stanescu}}]{DasSarma_PRB2012}%
  \BibitemOpen
  \bibfield  {author} {\bibinfo {author} {\bibfnamefont {S.}~\bibnamefont
  {Das~Sarma}}, \bibinfo {author} {\bibfnamefont {J.~D.}\ \bibnamefont {Sau}},
  \ and\ \bibinfo {author} {\bibfnamefont {T.~D.}\ \bibnamefont {Stanescu}},\
  }\href {\doibase 10.1103/PhysRevB.86.220506} {\bibfield  {journal} {\bibinfo
  {journal} {Phys. Rev. B}\ }\textbf {\bibinfo {volume} {86}},\ \bibinfo
  {pages} {220506} (\bibinfo {year} {2012})}\BibitemShut {NoStop}%
\bibitem [{\citenamefont {Brinkman}\ \emph {et~al.}(1970)\citenamefont
  {Brinkman}, \citenamefont {Dynes},\ and\ \citenamefont
  {Rowell}}]{Brinkman_JAP1970}%
  \BibitemOpen
  \bibfield  {author} {\bibinfo {author} {\bibfnamefont {W.~F.}\ \bibnamefont
  {Brinkman}}, \bibinfo {author} {\bibfnamefont {R.~C.}\ \bibnamefont {Dynes}},
  \ and\ \bibinfo {author} {\bibfnamefont {J.~M.}\ \bibnamefont {Rowell}},\
  }\href {http://link.aip.org/link/?JAP/41/1915/1} {\bibfield  {journal}
  {\bibinfo  {journal} {J. Appl. Phys.}\ }\textbf {\bibinfo {volume} {41}},\
  \bibinfo {pages} {1915} (\bibinfo {year} {1970})}\BibitemShut {NoStop}%
\end{thebibliography}
%

\end{document}